\documentclass[12pt]{article}

\usepackage[a4paper,margin=2.5cm]{geometry}
\usepackage{graphicx}
\usepackage{subfig}
\usepackage{subfloat}
\usepackage{dcolumn}
\usepackage{bm}

\usepackage[utf8]{inputenc}
\usepackage[T1]{fontenc}
\usepackage{mathptmx}
\usepackage{amsmath}
\usepackage{etoolbox}
\usepackage[authoryear,round]{natbib}
\usepackage{hyperref}
\usepackage{xcolor}

\title{Effective One-Body Interactions Due To The Presence Of A Liquid-Vapor
Interface}

\author{Melih G\"ul\\
\texttt{melih.guel@uni-tuebingen.de}
\and
Roland Roth\\
\texttt{roland.roth@uni-tuebingen.de}\\[0.5em]
\normalsize Institute for Theoretical Physics, University of T\"ubingen,\\
\normalsize Auf der Morgenstelle 14, 72076 T\"ubingen, Germany}

\date{\today}

\begin{document}

\maketitle

\begin{abstract}
In this study we investigate the behavior of additive binary square-well
mixtures within the framework of classical density functional theory. By
leveraging on the geometrical structure of the square-well interaction, we 
propose a novel form of the perturbation theory contribution to the density
functional in terms of weighted densities, inspired by fundamental measure
theory. We apply this functional in order to study the effective one-body
interaction due to the liquid-vapor interface of the solvent acting on a 
dilute component of dissolved nano particles. The effective one-body 
interaction attracts the nano particles strongly to the interface. We show
that this effective interaction potential can be calculated either from the
density profiles of the full mixture, at low but non-vanishing concentrations
of the nano particles, or by employing the Widom insertion theorem in the
dilute limit of vanishing density of nano particles. Both routes display
excellent agreement.
\end{abstract}




\section{Introduction}

In fluid mixtures, especially in those with particles of different
sizes, it is possible and often beneficial to map the full mixture
onto an effective one-component system by integrating out the
degrees of freedom of the other components \cite{McMillan45,Dijkstra98,Dijkstra99a,Dijkstra99b,Roth01,Hansen06}. Particles of the 
remaining effective one-component system interact via {\em effective}
interactions \cite{Likos01} that can result in a rich behavior. Well known and studied
examples for effective interactions in ideal colloid-polymer mixtures
are the entropic Asakura-Oosawa (AO) depletion interaction \cite{AO54,AO58,Lekkerkerker11} that induce an
effective attraction between colloids, despite the fact that all the
bare interaction in the model system are purely repulsive. If the size 
ratio between colloids with hard-sphere diameter $\sigma_c$ and polymer
with diameter $\sigma_p$ is sufficiently small, $\sigma_c/\sigma_p<0.1547$ \cite{Dijkstra99c},
the resulting AO depletion potential is a pair potential. For larger
polymer additional higher-body interactions emerge. An extension
of this idea to hard-sphere mixtures \cite{Roth00,Ashton11} is more complicated due to the
fact that the smaller spheres possess correlations and so the structure
of the resulting depletion potential is richer, with attractive and
repulsive parts. The correlations of the depletants also lead to effective
many-body contributions in the depletion potential, although the
two-body term is typically dominant and three- and higher-body terms
are neglected.

While the effective depletion interaction in an ideal colloid-polymer
mixture can be calculated from purely geometric arguments, i.e. by calculating
the overlap of excluded volume \cite{AO54,AO58,Lekkerkerker11}, already for hard-sphere mixtures the
problem is theoretically much more challenging \cite{Roth01,Ashton11}. An efficient and general theory
of effective interactions builds on Widom's insertion theorem \cite{Widom63,Henderson83}, that relates
the partition sum or the corresponding thermodynamic potential of a system
with an additional particle inserted into a inhomogeneous system to 
quantities {\em without} this particle. This idea was successfully employed
within the framework of classical density function theory (cDFT) \cite{Evans79,Evans92,Evans16} in the case
of hard-sphere mixtures. However, the theory does not rely on the fact
that hard spheres were considered but is valid for all inter-particle 
interactions. The description of fluid mixtures requires a reliable theory. In the case of hard-sphere mixtures, fundamental measure theory (FMT) \cite{Rosenfeld89,Roth10} provides
a very accurate framework. If more complicated interactions are present, a system of hard-sphere mixtures can be used as a good reference system for a perturbation theory treatment.

An important point that we would like to emphasize here is that the 
calculation of the effective pair interaction in a fluid mixture requires
one to take the dilute limit of the remaining component, typically the larger
one, in order to avoid to count the effect of particles of the remaining
effective one-component system multiple times. This allows one to split the
calculation of the effective pair interaction between two particles of the
remaining component into two parts. In the first part one calculates the
(inhomogeneous) density profile of the depletants in the presence of one
fixed particle of the remaining component. This is the so-called test-particle
geometry, where the {\em test particle} is fixed at a given position and hence
turned into an external potential for the depletants. In this calculation
the dilute limit for the remaining component implies that except for the fixed
test particle, that enters the calculation only in form of an external 
potential, no other particle of the remaining component has to be considered.
The second part of the calculation employs the Widom insertion theorem \cite{Widom63,Henderson83,Roth00} and
computes the change of the thermodynamic potential, which in the case of
DFT is the grand potential, due to the insertion of one particle of the
remaining component at a position relative to the fixed test particle. In order
to ensure that the resulting effective interaction potential vanishes at
infinite separation one subtracts from the change in the grand potential
due to the insertion the same quantity at infinite separation. A possible
interpretation of this calculation using Widom's insertion theorem \cite{Widom63,Henderson83} is that
we start with two particles of the remaining component, which are infinitely
separated and hence do not interact with each other. One of the particles is
removed from the system, i.e. is moved into the particle reservoir, and then
inserted into the system close to the fixed test particle \cite{Roth00}.

The situation that we aim to describe here is slightly different from the previous scenario. We wish to
calculate the effective interaction felt by solute particles due to the presence of a liquid-vapor interface formed by the
solvent, that we want to integrate out. One implication of this difference is that instead of a fixed test particle, the inhomogeneity originates from the existence of an interface. However, we still can employ Widom's insertion theorem \cite{Widom63,Henderson83} in order to calculate the effective interaction due to the interface. To this end we assume that one particle of the solute is infinitely far away from the interface, either in the homogeneous liquid or vapor phase. We then remove this particle from the system, or move it into the particle reservoir, and insert it in the vicinity of the liquid-vapor interface. The total change in
the grand potential due to these two steps, the removal and the insertion
of the particle, defines the effective interaction acting on particles of
the remaining component due to the presence of an interface \cite{Roth00}.

The phenomenon of phase separation in fluids, characterized by distinct 
liquid and vapor phases separated by an interface, is fundamental to
understanding the behavior of many systems. For any system consisting of a
solvent that phase separates and dissolved nano particles, effective 
interactions acting on the nano particles are generically most pronounced 
close to the liquid-vapor interface. A study in \cite{AbeKoga2014} investigated the behavior of solute particles by calculating their solubility near a liquid-vapor interface using Lennard-Jones and water-like solvents based on molecular dynamics simulation in the $NVT$ ensemble. They observed an enhancement of solubility near the interface due to attractive forces between solute and solvent particles. In \cite{Wise2018} the interfacial adsorption in water droplets is described as a competition between entropic and energetic contributions of the solute-water interaction. Further studies on the interfacial gas adsorption of real fluids were carried out in \cite{Vergara2019} employing weighted density approximations (WDA) within DFT and in \cite{Sauer2017} a similar study was performed using WDA for fluid-liquid interfaces of mixtures. Both studies found a good agreement in calculating the interfacial surface tension of the mixture due to the presence of spherical molecules at the interface. The work on bubble nucleation \cite{Talanquer2001} using Lennard-Jones potential with random phase approximation (RPA) showed adsorption of nitrogen or carbon dioxide and water mixture. Finally, in \cite{Okamoto2016} DFT was also applied to dilute binary mixtures with liquid-vapor phase separation in the form of RPA for attractive forces. Therein, density profiles of solvent and solute are calculated for a planar interface with the observation that solute particles lead to an enhanced adsorption at the interface.

To explore these interactions in detail, we employ a model system of
the square-well fluid mixture, recognized for its simplicity and efficacy
in modeling such phenomena. The square-well interaction potential incorporates
the particle diameter, interaction strength and interaction range as
parameters. This model serves as an illustrative framework for capturing 
the essential physics of the effect of a phase separating solvent on the
effective interaction of nano particles.

The manuscript is organized as follows. In Sec.~\ref{sec:theory} we introduce
the DFT for our model system. While in principle the DFT for square-well
fluids and fluid mixture within random phase approximation perturbation theory
\cite{Hansen06,Archer17}
is well established, we introduce an efficient way of treating the square-well
attraction by considering additive interactions. We provide new weighted densities in form of geometrical measures of the densities adjusted to square-well interactions within the framework of FMT permitting a systematic approach for square-well mixtures with an arbitrary number of components. In Sec.~\ref{sec:results}
we present our results, starting with bulk considerations for a binary
square-well mixture of a solvent and added nano particles at a low 
concentration. We investigate the effect of the nano particles on the phase
equilibrium between a liquid and a vapor phase and present the resulting density
profiles of the free interface for such a binary mixture. We analyze the
density profile of the nano particles also in the context of effective
interaction and show that in the dilute limit of vanishing concentration of
the nano particles, the effective interaction of the interface acting on
nano particles can be computed efficiently using Widom's insertion theorem \cite{Widom63,Henderson83}.
In Sec.~\ref{sec:summary} we summarize our findings and conclude with an
outlook.

\section{Theory} \label{sec:theory}

We study the behavior of a square-well mixture within the framework of
classical density functional theory \cite{Evans79,Evans92,Evans16} which is a
very powerful and
versatile framework to describe and predict both the inhomogeneous
structure and the corresponding thermodynamics of a many body system
subjected to an external potential. One can prove for a $\nu$-component mixture
that a functional of the grand potential exists and has the form
\begin{equation}\label{eq::omega}
	\Omega[\{\rho_i(\mathbf{r})\}]={\cal F}_\text{id}[\{\rho_i(\mathbf{r})\}]+{\cal F}_\text{ex}[\{\rho_i(\mathbf{r})\}]+\sum_{i=1}^\nu\int d\mathbf{r}\,\rho_i(\mathbf{r})\left(V^i_\text{ext}(\mathbf{r})-\mu^i\right),
\end{equation}
where ${\cal F}_\text{id}$ is the exactly known free energy of an ideal gas
\begin{equation}\label{eq::id-functional}
    \beta{\cal F}_\text{id}[\{\rho_i(\mathbf{r})\}] = \sum_{i=1}^\nu\int d\mathbf{r}\,\rho_i(\mathbf{r})\left(\log(\Lambda_i^3\rho_i(\mathbf{r}))-1\right),
\end{equation}
with $\Lambda_i$ the thermal wave length of species $i$.
${\cal F}_\text{ex}$ is the excess (over the ideal gas) intrinsic Helmholtz
free energy, that account for the interactions between particles, and 
$\mu^i$ is the chemical potential of species $i$. The external potential
$V^i_\text{ext}$ will be not considered in our study and is set to zero.
The functional of the excess free energy can be split
\begin{equation}\label{eq::fex}
{\cal F}_\text{ex}[\{\rho_i(\mathbf{r})\}]={\cal F}_\text{hs}[\{\rho_i(\mathbf{r})\}]+{\cal F}_\text{sw}[\{\rho_i(\mathbf{r})\}]
\end{equation}
into a hard-sphere reference part ${\cal F}_\text{hs}$, that we treat within the framework of
fundamental measure theory (FMT) \cite{Rosenfeld89,Roth10}, and a perturbation contribution ${\cal F}_\text{sw}$ that
accounts for the square-well attraction within the (optimized) random
phase approximation \cite{Hansen06,Archer17}.

It can be shown that the functional of the grand potential, 
Eq.~(\ref{eq::omega}), is minimized by the equilibrium density profiles
$\rho_{i,0}(\mathbf{r})$
\begin{equation}\label{eq::ELG}
	\frac{\delta\Omega[\{\rho_i\}]}{\delta\rho_i(\mathbf{r})}\bigg|_{\rho_i(\mathbf{r})=\rho_{i,0}(\mathbf{r})}=0,
\end{equation}
and its value at the minimum reduces to the grand potential of the system.

The reference system of a hard-sphere mixture can be accurately described 
by the White-Bear version of FMT \cite{Roth02,YuWu02} which has the form
\begin{equation}\label{eq::WB-Functional}
	\beta{\cal F}_\text{hs}[\{\rho_i(\mathbf{r})\}] = \int d\mathbf{r}\,\Phi_\text{hs}(\{n_\alpha\}),
\end{equation}
where the free energy density $\Phi_\text{hs}$ is a function of weighted densities 
\begin{equation}\label{eq::hs-weighted-densities}
    n_\alpha(\mathbf{r}) = \sum_{i=1}^\nu\int d\mathbf{r}'\,\rho_i(\mathbf{r}') w^i_\alpha(\mathbf{r}-\mathbf{r}'),
\end{equation}
with the sum running over all $\nu$ components of the hard-sphere mixture. 
The weight functions $w^i_\alpha(\mathbf{r})$ characterize the size and shape
of the hard spheres.

From Eq.\eqref{eq::ELG} we obtain an implicit equation for the equilibrium
density profile
\begin{equation}\label{eq::implicit-equation}
\rho_i(\mathbf{r})=\rho_i\exp\left(c^{(1)}_i(\mathbf{r})+\beta\mu^i_\text{ex}-\beta V^i_\text{ext}(\mathbf{r})\right).
\end{equation}
Here, $\rho_i$ is the bulk density, $c^{(1)}_i(\mathbf{r})$ is the one-body 
correlation function and $\mu^i_{ex}$ is the excess chemical potential of 
component $i$ which in this case is given by the sum of hard-sphere and SW 
contributions  $\mu^i_\text{ex}=\mu^i_\text{hs}+\mu^i_\text{sw}$.
Eq.\eqref{eq::implicit-equation} cannot be solved analytically since the
density profile $\rho({\bf r})$ also appears on the right-hand side,
contributing to the one-body correlation function $c^{(1)}_i(z)$.
Numerical schemes, like the Picard iteration, allow us to solve
Eq.\eqref{eq::implicit-equation}.

The remaining part of the functional is the perturbation treatment of
the square-well interaction, which we will discuss in some details, assuming
additive mixtures.
\subsection{Square-well interaction within DFT}

The square-well interaction (SW) is a commonly used model to realize 
liquid-vapor coexistence with a simple interaction potential that possesses both
a hard-core repulsion at short distances preventing particle to overlap, and a 
short-ranged attraction. The interaction potential takes the form
\begin{equation}\label{eq::SW-potential}
	\phi^{(i,j)}(r) = \begin{cases}
            \infty,\quad r\leq\sigma_{ij}\\
		-\epsilon_{ij},\quad \sigma_{ij}<r<\lambda_{ij}\sigma_{ij},\\
		0,\quad\text{otherwise},
	\end{cases}
\end{equation} 
where $\sigma_{ij}=(\sigma_i+\sigma_j)/2$ are the distances of closest
approach. Here $\sigma_i$ is the hard core diameter of species $i$, 
$\epsilon_{ij}$ is the interaction strength and $\lambda_{ij}$ is the 
interaction range. The hard core repulsion is taken into account by the
hard-sphere reference system, and the attractive square-well part is 
treated within the well-known (optimized) random-phase approximation (RPA)
\cite{Hansen06,Archer17}
\begin{equation}\label{eq::SW-Functional}
	\beta{\cal F}_\text{sw}[\{\rho_i(\mathbf{r})\}] = \sum_{i,j}^\nu\frac{1}{2}\int d\mathbf{r}\int d\mathbf{r}'\,\rho_i(\mathbf{r})\rho_j(\mathbf{r}')\beta\phi^{(i,j)}_\text{sw}(\mathbf{r}-\mathbf{r}'),
\end{equation}
where the SW potential $\phi^{(i,j)}_\text{sw}(\mathbf{r})$ is extended inside 
the hard core 
\begin{equation}\label{eq::sw-potential-ij}
    \phi^{(i,j)}_\text{sw}(\mathbf{r})=
    \begin{cases}
        -\epsilon_{ij},\quad 0<r<\lambda_{ij}\sigma_{ij}\\
        0,\quad \text{otherwise}.
    \end{cases} ,
\end{equation}
in order to compensate the lack of correlations in the interaction term.

The total bulk free energy density $f$ is obtained by evaluating at bulk densities $\rho_i(\mathbf{r})\rightarrow\rho_i$
\begin{equation}\label{eq::ftot-bulk}
    f =\frac{{\cal F}_\text{id}[\{\rho_i\}]+{\cal F}_\text{hs}[\{\rho_i\}]+{\cal F}_\text{sw}[\{\rho_i\}]}{V} 
\end{equation}
and particularly for the SW expression we find
\begin{equation}\label{eq::fsw-bulk}
    f_\text{sw} = \frac{{\cal F}_\text{sw}[\{\rho_i\}]}{V}= -\frac{2\pi}{3}\sum^\nu_{i,j}\rho_i\rho_j(\lambda_{ij}\sigma_{ij})^3\epsilon_{ij}.
\end{equation}
From Eq.\eqref{eq::fsw-bulk} we can derive the square-well contributions to 
the chemical potential $\mu^i_\text{sw}$ and the pressure $P_\text{sw}$ as
\begin{equation}\label{eq::sw-chempot}
    \mu^i_\text{sw} = \frac{\partial f_\text{sw}}{\partial \rho_i} = -\frac{4\pi}{3}\sum_{j=1}^\nu\rho_j(\lambda_{ij}\sigma_{ij})^3\epsilon_{ij},
\end{equation}
and
\begin{equation}\label{eq::sw-pressure}
    P_\text{sw} = -f_\text{sw}+\sum_{i=1}^\nu\rho_i\mu^i_\text{sw} = -\frac{2\pi} {3}\sum^\nu_{i,j}\rho_i\rho_j(\lambda_{ij}\sigma_{ij})^3\epsilon_{ij}.
\end{equation}

\subsection{New Functional for SW Mixtures}
Besides the usual expression Eq.\eqref{eq::SW-Functional}, which can be
cumbersome if a mixture of several components is considered, we want to make
use of the geometrical structure of the square-well interaction.
Inspired by the low-density expression for a hard-sphere system
\begin{equation}\label{low-dens-functional}
	\lim_{\rho_i\to0}\beta{\cal F}_\text{hs}[\{\rho_i(\mathbf{r})\}]=\int d\mathbf{r}\,\left(n_0(\mathbf{r})n_3(\mathbf{r})+n_1(\mathbf{r})n_2(\mathbf{r})-\mathbf{n}_1(\mathbf{r})\cdot\mathbf{n}_2(\mathbf{r})\right),
\end{equation}
where the r.h.s. is the {\em exact} low density limit written in terms of
weighted densities of FMT \cite{Rosenfeld89,Roth10}. Since the square-well perturbation term within the
optimized RPA has the same geometrical structure as the hard-sphere low-
density limit, it is possible to define weight functions
$w^\epsilon_\alpha$ of the SW fluid in such a way that we get a
functional equivalent to Eq.\eqref{eq::SW-Functional}. By defining
\begin{gather}\label{eq::SW-weight-functions}
	w^{\epsilon,i}_3(\mathbf{r})=\sqrt{\epsilon_i}\,\Theta(\lambda_i R_i-|\mathbf{r}|),\quad
	w^{\epsilon,i}_2(\mathbf{r}) = \sqrt{\epsilon_i}\,\delta(\lambda_i R_i-|\mathbf{r}|),\quad
	\mathbf{w}^{\epsilon,i}_2(\mathbf{r}) = w^{\epsilon,i}_2(\mathbf{r})\frac{\mathbf{r}}{r}\\\nonumber
	w^{\epsilon,i}_1(\mathbf{r}) = -\frac{w^{\epsilon,i}_2(\mathbf{r})}{4\pi \lambda_i R_i},\quad w^{\epsilon,i}_0(\mathbf{r}) = -\frac{w^{\epsilon,i}_2(\mathbf{r})}{4\pi (\lambda_i R_i)^2},\quad \mathbf{w}^{\epsilon,i}_1(\mathbf{r}) = -\frac{\mathbf{w}^{\epsilon,i}_2(\mathbf{r})}{4\pi\lambda_i R_i}
\end{gather}
we obtain
\begin{equation}\label{eq::SW-functional-alternative}
	\beta{\cal F}_\text{sw}[\{\rho_i(\mathbf{r})\}]=\int d\mathbf{r}\,\beta\left(n^\epsilon_0(\mathbf{r})n^\epsilon_3(\mathbf{r})+n^\epsilon_1(\mathbf{r})n^\epsilon_2(\mathbf{r})-\mathbf{n}^\epsilon_1(\mathbf{r})\cdot\mathbf{n}^\epsilon_2(\mathbf{r})\right)
\end{equation}
which is  the RPA of the SW fluid, now in the fashion of FMT. It should be
noticed that in order to produce the correct sign in the case of SW
interaction, the weight functions $w^{\epsilon,i}_2(\mathbf{r})$, $w^{\epsilon,i}_1(\mathbf{r})$ and $\mathbf{w}^{\epsilon,i}_1(\mathbf{r})$
must have a minus sign. For square-shoulder (SS) interactions, these minus 
signs would be absent.

Similar to Eq.\eqref{eq::hs-weighted-densities} the weighted densities $n^\epsilon_\alpha(\mathbf{r})$ are given by
\begin{equation}\label{eq::sw-weighted-densities}
    n^\epsilon_\alpha(\mathbf{r}) = \sum_{i=1}^\nu\int d\mathbf{r}'\,\rho_i(\mathbf{r}) w^{\epsilon,i}_\alpha(\mathbf{r}-\mathbf{r}')
\end{equation}
which attain the following values in the bulk
\begin{gather}\label{eq::sw-weighted-densities-bulk}
    n^\epsilon_3(\mathbf{r})\rightarrow\frac{4\pi}{3}\sum_{i=1}^\nu(\lambda_i R_i)^3\rho_i\sqrt{\epsilon_i},\quad n^\epsilon_2(\mathbf{r})\rightarrow4\pi\sum_{i=1}^\nu (\lambda_i R_i)^2\rho_i\sqrt{\epsilon_i},\\\nonumber
    n^\epsilon_1(\mathbf{r})\rightarrow-\sum_{i=1}^\nu \lambda_i R_i\rho_i\sqrt{\epsilon_i},\quad n^\epsilon_0(\mathbf{r})\rightarrow-\sum_{i=1}^\nu \rho_i\sqrt{\epsilon_i},\\\nonumber
    \mathbf{n}^\epsilon_2(\mathbf{r})\rightarrow 0,\quad \mathbf{n}^\epsilon_1(\mathbf{r})\rightarrow 0.
\end{gather}
Then, the excess free energy density $f_\text{sw}$ of the SW mixture is found to be
\begin{gather}\label{eq::fsw-mixing}
    f_\text{sw} = -\frac{2\pi}{3}\sum_{i=1}^\nu \rho_i^2(\lambda_i \sigma_i)^3\epsilon_i-\frac{\pi}{6}\sum_{(i,j)}^\nu \rho_i\rho_j\left(\lambda_i\sigma_i+\lambda_j\sigma_j\right)^3\sqrt{\epsilon_i\epsilon_j},
\end{gather}
where the first term on the r.h.s. refers to SW interaction of component $i$ with itself and equals those of Eq.\eqref{eq::fsw-bulk}. The second term of Eq.\eqref{eq::fsw-mixing}, which it is summed over all possible pairs $(i,j)$ with $\,i\neq j$, accounts for inter-component SW interactions wherein the energies and ranges are predefined by the energies and ranges of the components, respectively.

For a one-component SW fluid this FMT inspired approach seems to be more
complicated, however, by considering a multi-component mixture the simplicity 
of Eq.\eqref{eq::SW-functional-alternative} becomes apparent. As for the 
multi-component hard-sphere mixture, also here the concise description of 
the excess SW functional is advantageous. We want to emphasize also that the
mixing rule according to Lorentz-Berthelot is already inherent in this
description, i.e. given the SW energies $\epsilon_i$ and ranges $\lambda_i$,
the inter-component energies and ranges are fixed. For example, the energy
$\epsilon_{ij}$ between component $i$ and $j$ is given by
\begin{equation}\label{eq::energy-mixing-rule}
	\epsilon_{ij}=\sqrt{\epsilon_i\epsilon_j}
\end{equation}
with the range
\begin{equation}\label{eq::range-mixing-rule}
	\lambda_{ij}=\frac{\lambda_i \sigma_i+\lambda_j \sigma_j}{\sigma_i+\sigma_j}.
\end{equation}
Imposing these mixing rules, Eq.\eqref{eq::energy-mixing-rule} and 
Eq.\eqref{eq::range-mixing-rule}, onto the SW free energy density 
Eq.\eqref{eq::fsw-bulk} in the most general case, shows the equivalence to 
Eq.\eqref{eq::fsw-mixing} hence describing the same thermodynamics. 

We see that in the Lorentz-Berthelot mixing rule the inter-component energy $\epsilon_{ij}$ is the geometric mean of $\epsilon_i$ and $\epsilon_j$, i.e. being smaller or equal than the maximum of these two. This is a restriction that can prohibit configurations where a component $i$ prefers to stay with a different component $j$ rather than with itself if $\epsilon_i>\epsilon_j$. Thus, it is still necessary to include the case of an inter-component energy $\epsilon_{ij}$ which is independent.

\section{Results} \label{sec:results}

\subsection{Density Profiles}

Here we focus on the case $\nu=2$ where we study the phase coexistence of the binary SW mixture using the free energy density in Eq.\eqref{eq::fsw-mixing}. Therefore, we have three contributions to $f_\text{sw}$
\begin{gather}\label{eq::fsw-binary}
    f_\text{sw}(\rho_1,\rho_2) = -\frac{2\pi}{3}\lambda_1^3\sigma_1^3\rho_1^2\epsilon_1-\frac{2\pi}{3}\lambda_2^3\sigma_2^3\rho_2^2\epsilon_2\\\nonumber
    -\frac{\pi}{6}(\lambda_1\sigma_1+\lambda_2\sigma_2)^3\rho_1\rho_2\sqrt{\epsilon_1\epsilon_2}.
\end{gather}
Here, the first two contributions are assigned to SW interactions of the two components, and the third contribution to the SW interaction between the first and the second component.

The total pressure $P(\rho_1,\rho_2)$ of the binary SW mixture reads, assuming
the White-Bear version of FMT \cite{Roth02,YuWu02}, as
\begin{equation}\label{eq::pressure-binary}
    P(\rho_1,\rho_2) = P_\text{BMCSL}(\rho_1,\rho_2)+P_\text{sw}(\rho_1,\rho_2)
\end{equation}
in which $P_\text{BMCSL}$ refers to the Boublik-Mansoori-Carnahan-Starling-Leland (BMCSL) \cite{MCSL03} pressure, a generalization to  hard-sphere mixtures of the Carnahan-Starling expression for pure fluids \cite{CS69}, expressed in terms of scaled-particle variables \cite{Reiss04,SPT04} which are identified by the bulk limits of the weighted densities, Eq.\eqref{eq::hs-weighted-densities}. The pressure $P_\text{sw}$, Eq.~(\ref{eq::sw-pressure}), due to SW interactions is derived from the free energy density Eq.\eqref{eq::fsw-binary}. The chemical potentials $\mu_1(\rho_1,\rho_2)$ and $\mu_2(\rho_1,\rho_2)$ are obtained in a similar way and are given in Eq.~(\ref{eq::sw-chempot}).

Since we are interested in the effective interaction of a liquid-vapor
interface of a solvent (component 1) on dissolved nano particles (component 2),
we first have to establish equilibrium conditions for a binary mixture.
The conditions for phase coexistence are given by
\begin{align}\label{eq::phasecoex-binary}
    P(\rho_1^l,\rho_2^l) &= P(\rho_1^v,\rho_2^v)\\\nonumber
    \mu_1(\rho_1^l,\rho_2^l) &= \mu_1(\rho_1^v,\rho_2^v)\\\nonumber
    \mu_2(\rho_1^l,\rho_2^l) &= \mu_2(\rho_1^v,\rho_2^v)
\end{align}
which are fulfilled at a given temperature $T$. The densities $\rho^l_i$ and $\rho^v_i$ refer to the liquid ($l$) and the vapor ($v$) density of component $i$, respectively.

We see that Eq.\eqref{eq::phasecoex-binary} provides us with three equations 
for four unknowns $\rho^l_i$ and $\rho^v_i$ with $i=1,2$. Hence, we further 
provide a fixed low value for the liquid density 
$\rho^l_2=\rho_{2,0}=10^{-6}\times 6/(\pi\sigma_2^3)$ and solve 
Eq.\eqref{eq::phasecoex-binary} for the remaining densities given the SW 
parameters $\epsilon_i$ and $\lambda_i$. 

We regard the first component 
as solvent with fixed SW parameters $\beta\epsilon_1=1$ and 
$\lambda_1=1.5$, and the second component as dilute solute composed of nano
particles, i.e. $\rho_2\ll\rho_1$ has to be satisfied. Then, we expect 
the solute not to considerably change the phase behavior of the solvent, which specifically means that the coexisting densities $\rho^l_1$ 
and $\rho^v_1$ of the pure one-component SW fluid remain virtually unchanged 
with values
\begin{equation}\label{eq::coex-densities-one-comp}
    \rho^l_1\sigma_1^3 = 0.6016,\quad \rho^v_1\sigma_1^3 = 0.0370
\end{equation}
and a reduced interfacial surface tension 
\begin{equation}\label{eq::surface-tension-one-comp}
    \beta\gamma\sigma_1^2 = 0.20906 
\end{equation}

\begin{figure}
\centering
\includegraphics[scale=0.8]{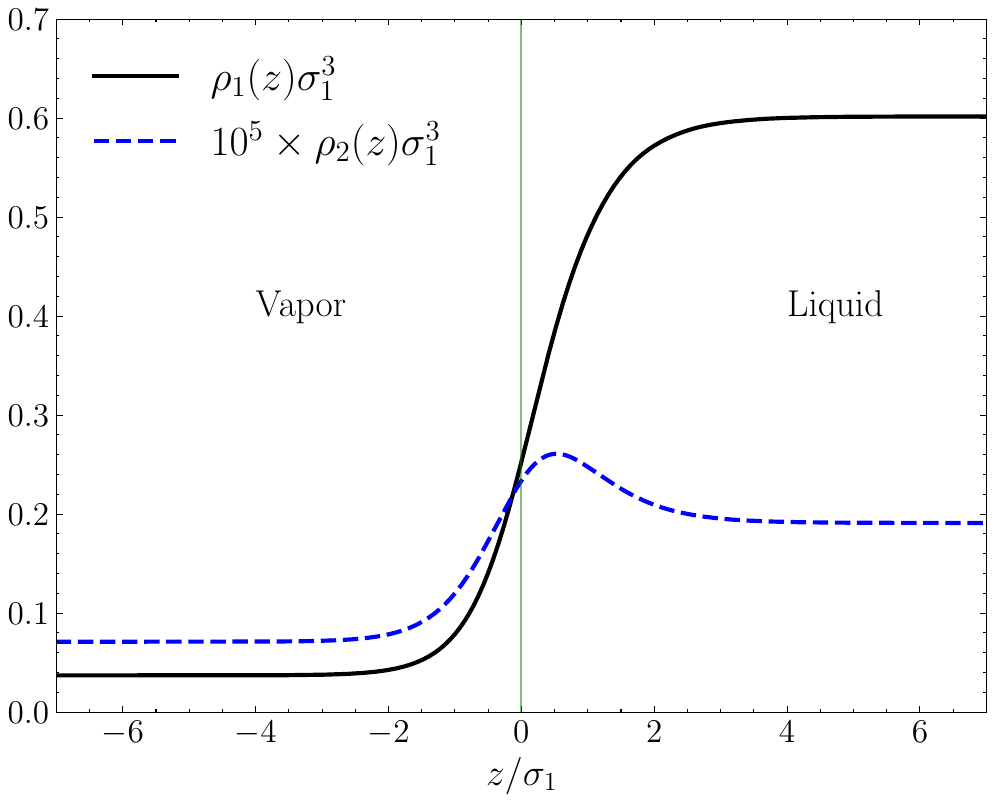}
\caption{Liquid-vapor phase separation of the SW binary mixture with corresponding interface (green vertical line) located at $z=0$ and density profiles of the first component (black solid) and the second component (dashed blue) which we magnified for visualization. The SW parameters of the second component are chosen to be $\beta\epsilon_2=0.6$ and $\lambda_2=1.5$ with $\sigma_2/\sigma_1=1$.\label{fig::liquid_vapor}}
\end{figure}

We solve Eq.\eqref{eq::phasecoex-binary} numerically for several values of the
interaction parameter $\epsilon_2$ and size ratio $\sigma_2/\sigma_1$. We
further perform a 
minimization with cDFT according to Eq.\eqref{eq::implicit-equation} employing 
the functional proposed in Eq.\eqref{eq::SW-functional-alternative} in order to 
obtain density profiles $\rho_1(z)$ and $\rho_2(z)$ for the planar free
interface between the liquid and vapor phases. The interaction range $\lambda_2$ of the solute is set such that the corresponding width $\sigma_2(\lambda_2-1)$ 
is equal to the width $\sigma_1(\lambda_1-1)$ of the first component, i.e. we
have $\lambda_2 = 1+(\lambda_1-1)\sigma_1/\sigma_2$.
Particularly, the symmetric case where $\rho^l_2=\rho^v_2$ is of interest and is realized for a specific interaction energy $\epsilon_2$ of the second component.

The qualitative behavior of the solubility can be well understood considering the equilibrium condition of the grand canonical potential density $\omega(\rho_1,\rho_2)=\Omega(\rho_1,\rho_2)/V=f(\rho_1,\rho_2)-\mu_1\rho_1-\mu_2\rho_2$ in the dilute limit $\rho_2\rightarrow 0$. Hence, the equilibrium distribution of the solvent $\rho_1(z)$ is assumed to be unaffected by the nano particles and can be described in a good approximation as a sigmoidal of the form
\begin{equation}\label{eq::rho1-sigmoid}
    \rho_1(z)=\rho_1^v + \frac{\rho_1^l-\rho_1^v}{1+\exp(-az)},
\end{equation}
where $a\approx1.79/\sigma_1$ is obtained from a fit using the SW values of the solvent, $\beta\epsilon_1=1.0$ and $\lambda_1=1.5$. At equilibrium, we have $\partial\omega/\partial\rho_2=0$ from which in leading order
\begin{equation}\label{eq::omega-min}
    \log\left(\frac{\rho_2(z)}{\rho_{2,0}}\right)+\beta V_\text{eff}(\rho_1(z),\epsilon_{12},\lambda_{12})=0
\end{equation}
follows, with an effective potential
\begin{equation}\label{eq::veff}
    V_\text{eff}(\rho_1(z),\epsilon_{12},\lambda_{12})=\frac{\partial \Delta f(\rho_1,\rho_2)}{\partial\rho_2}\bigg|_{\rho_2=0}.
\end{equation}
Here $\Delta f(\rho_1,\rho_2)$ is the change in free energy due to the insertion of solutes. Note that due to the linearization with respect to $\rho_2$, only the inter-component interaction parameters $\epsilon_{12}$ and $\lambda_{12}$ contribute in Eq.\eqref{eq::veff}. In \cite{AbeKoga2014}, an equivalent investigation to Eq.\eqref{eq::omega-min} was done by applying the potential distribution theorem providing a description of the excess adsorption of solute particles at the liquid-vapor interface. The solubility thus is connected to the local excess chemical potential of the solute particle. Abe and Koga showed that varying the interaction energy leads to different shapes of solubility that can exhibit local maxima at the interface. Finally, Eq.\eqref{eq::omega-min} provides an effective description of a system only being comprised of solute particles, see Sec.\ref{sc::effective-interactions}.

In Fig.\ref{fig::liquid_vapor} we present one scenario of the binary SW mixture 
where both components prefer to stay together (good solvent) and we divide the
system after phase separation into vapor ($z<0$) and liquid ($z>0$). We see
that the solvent assumes its coexisting densities of vapor and liquid 
sufficiently far away from the interface located at $z=0$ where it makes a
transition, that is well described by a sigmoidal function. According to our
assumption that $\rho_2\ll\rho_1$, the coexisting densities and the shape of
the density profile are only slightly affected by the presence of the solute. 
We observe an adsorption of nano particles at the interface with a pronounced
maximum.

\begin{figure}
\centering
\includegraphics[scale=0.8]{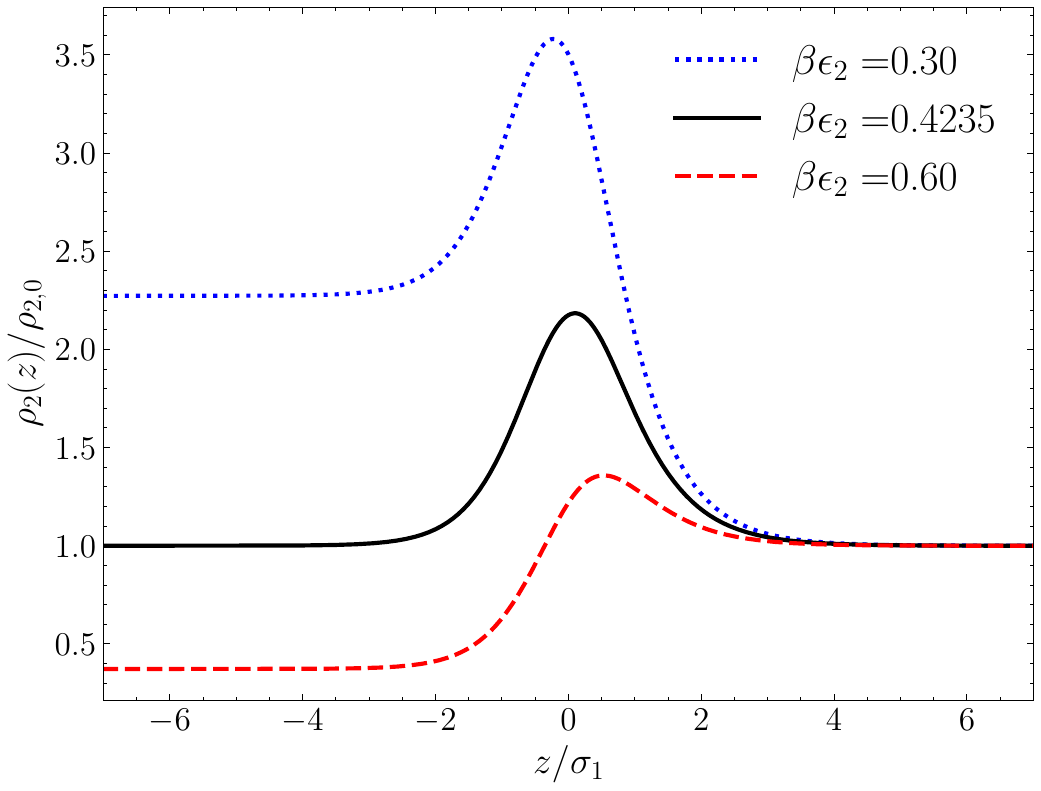}
\caption{Normalized density profile $\rho_2(z)/\rho_{2,0}$ of the second 
component at the interface $z=0$ for three energies $\epsilon_2$ with range 
$\lambda_2=1.5$ and size ratio $\sigma_2/\sigma_1=1$. For reasons of clarity
we do not show the density profile of the solvent. The lowest value of 
$\epsilon_2$ corresponds to the case of (slightly) solvophobic nano particles,
that prefer the vapor phase, while the highest value of $\epsilon_2$
corresponds to case of (slightly) solvophilic nano particles, which prefer
the liquid phase. In the case in between neither the liquid nor the vapor
phase is prefered by the nano particles.\label{fig::1_to_1}}
\end{figure}

We can alter the behavior of the solute with respect to the solvent by
changing the energy $\epsilon_2$. By doing that we change the interaction
energy $\epsilon_{12}$ according to Eq.\eqref{eq::energy-mixing-rule}.
Figure~\ref{fig::1_to_1} displays some density profiles of the nano particles 
normalized by $\rho_{2,0}$ and for energies $\epsilon_2$ located slightly below and above 
the energy $\beta\epsilon_2=0.4235$ for which the coexisting densities 
$\rho^l_2$ and $\rho^v_2$ are equal. The particles prefer neither the liquid 
nor the vapor phase thus gather at the interface. For reasons of clarity we do not
plot the interface profile of the solvent. All three density profiles show a maximum close to the interface. The nano particles are attracted
to the interface in the presence of (slightly) solvophobic and (slightly)
solvophilic nano particles. The maximum at the interface becomes more
pronounced as we lower the value of $\epsilon_2$. In addition, as we include
$\rho_2(z\rightarrow\infty)=\rho_{2,0}$ as a fourth equation to 
Eq.\eqref{eq::phasecoex-binary}, the normalized density profiles approach unity 
for $z\rightarrow\infty$.

\begin{figure}
\centering
\includegraphics[scale=0.6]{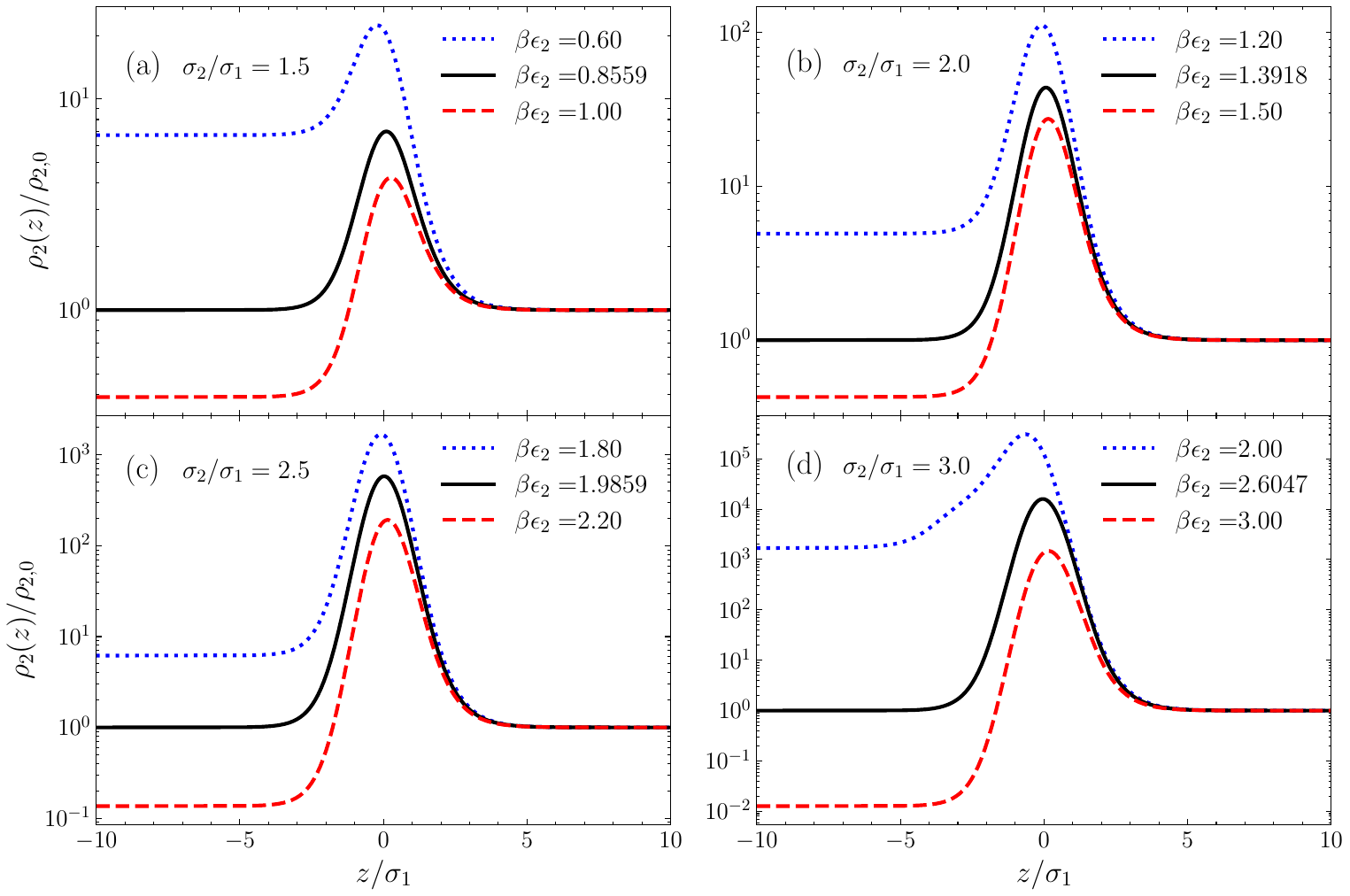}
\caption{Normalized density profiles $\rho_2(z)/\rho_{2,0}$ for several size ratios $\sigma_2/\sigma_1=1.5,\,2.0,\,2.5,\,3.0$ and energies $\epsilon_2$ slightly below (dotted blue) and above (dashed red) the symmetric solutions (solid black). By increasing the size ratio the adsorption at the interface becomes stronger and enhances the density by several orders of magnitude. Note the logarithmic scaling of the ordinates.\label{fig::total}}
\end{figure}

Now, we extend our discussion to size ratios $\sigma_2/\sigma_1>1$ as shown
in Fig.\ref{fig::total}. We again consider the density profile $\rho_2(z)$ 
for values of $\epsilon_2$  below and above
the symmetric case. Even in the slightly asymmetric mixture shown in Fig.\ref{fig::total}(b), we observe a more pronounced density maximum in $\rho_2(z)$ at the liquid-vapor interface, compared to the symmetric mixture shown in Fig.\ref{fig::1_to_1}. Furthermore, the energy $\epsilon_2$ required
for the symmetric solution is roughly doubled. By further increasing the size 
ratio the accumulation of nano
particles  at the interface becomes stronger, reaching around three orders of 
magnitude in Fig.\ref{fig::total}(c) and up to five orders of magnitude in 
Fig.\ref{fig::total}(d). Therefore, the resulting energy $\epsilon_2$ has to increase in order to keep the particles at the interface. By going to the highest asymmetric 
mixture $\sigma_2/\sigma_1=3$ considered here, Fig.\ref{fig::total}(d), we see 
some additional structure of the density profile $\rho_2(z)$ for 
$\beta\epsilon_2=2$ towards the vapor phase.

\begin{table}[h]
    \centering
    \scalebox{1.02}{
    \begin{tabular}{|c|c|c|c|c|c|}
        \hline
         $\sigma_2/\sigma_1$ & 1 & 1.5 & 2.0 & 2.5 & 3.0 \\\hline
        \textcolor{black}{DFT} & $-9.0\cdot 10^{-6}$ & $-1.0\cdot 10^{-5}$ & $-2.0\cdot 10^{-5}$ & $-9.3\cdot 10^{-5}$ & $-1.2\cdot 10^{-3}$\\ \hline
        \textcolor{black}{Gibbs} & $-1.3\cdot 10^{-6}$ & $-1.7\cdot 10^{-6}$ & $-4.0\cdot 10^{-6}$ & $-2.2\cdot 10^{-5}$ & $-3.0\cdot 10^{-4}$\\ \hline
    \end{tabular}
    }
    \caption{\textcolor{black}{Reduction $\tilde{\gamma}-\tilde{\gamma}_0$ in the surface tension $\tilde{\gamma}=\beta\gamma\sigma_1^2$ of the binary SW mixture calculated for the symmetric solution $\rho^l_2=\rho^v_2$ for different size ratios $\sigma_2/\sigma_1$ using DFT and  from the Gibbs adsorption theorem, Eq.\eqref{eq::gibbs-surface-tension}. Due to our assumption $\rho_2\ll\rho_1$, only a small deviation from the surface tension of the one-component system, Eq.\eqref{eq::surface-tension-one-comp}, is observed.}}
    \label{tab:surface_tension}
\end{table}

The addition of nano particles reduces the surface tension $\gamma$ of the
liquid-vapor interface as they accumulate at the interface which is 
significantly enhanced in the case of highly asymmetric mixtures, see 
Fig.\ref{fig::total}(c) and (d). This adsorption of particles lowers the free 
energy associated with the interface and consequently leads to a reduction of 
the surface tension. In Tab.\ref{tab:surface_tension} we present values of the 
reduced surface tension $\beta\gamma\sigma_1^2$ for several size ratios 
$\sigma_2/\sigma_1$. Since we only allow for a very low density reservoir 
$\rho_2$ of the second component, the expected reduction of the surface 
tension is very small, around 0.6\%. The change of the surface tension as we increase the size ratio $\sigma_2/\sigma_1$ from 2.5 to 3, that can be seen
in Tab.\ref{tab:surface_tension}, can be explained by the fact, that the
chemical potential $\mu_2$ and the excess adsorption $\Gamma_2$ increase with the size ratio. 
For a binary mixture the Gibbs 
adsorption theorem reads
\begin{equation}\label{eq::gibbs-ads-th}
    \text{d}\gamma = -\Gamma_1 \text{d}\mu_1-\Gamma_2 \text{d}\mu_2,
\end{equation}
where $\Gamma_i,\,i=1,2$ is the excess adsorption of component $i$ that depends on the location $z^*$ of the Gibbs dividing surface $A=\sigma_1^2$. We choose $z^*$ such that $\Gamma_1=0$, i.e.
\begin{equation}\label{eq::condition-gibbs-surface}
    0=\Gamma_1 = \frac{1}{A}\left(\int_{-\infty}^{z^*}\text{d}z\,(\rho_1(z)-\rho_1^v)+\int_{z^*}^{\infty}\text{d}z\,(\rho_1(z)-\rho_1^l)\right)
\end{equation}
is fulfilled by $z^*$. \textcolor{black}{In the same manner, the excess adsorption $\Gamma_2$ is obtained
\begin{equation}\label{eq::excess-adsorption-2}
    \Gamma_2 = \frac{\rho_{2,0}}{A}\int_{-\infty}^\infty \text{d}z\,\left(e^{-\beta V_\text{eff}(z)} - 1\right),
\end{equation}
where we used Eq.\eqref{eq::omega-min} and the symmetric case $\rho^l_{2}=\rho^v_2=\rho_{2,0}$.}
Then, from Eq.\eqref{eq::gibbs-ads-th} we obtain as a first approximation in the dilute limit
\textcolor{black}{
\begin{equation}\label{eq::gibbs-surface-tension}
    \beta\gamma \approx \beta\gamma_0-\Gamma_2,
\end{equation}
}
where $\gamma_0$ is the reference surface tension, Eq.\eqref{eq::surface-tension-one-comp}, of the one-component case. \textcolor{black}{At this point we can already infer from Eq.\eqref{eq::excess-adsorption-2} that the excess adsorption $\Gamma_2$ must be positive, since the integrand is positive due to $\beta V_\text{eff}(z)<0$. Thus, the surface tension $\gamma$, according to Eq.\eqref{eq::gibbs-surface-tension}, is reduced, in agreement with our expectation.}
We can therefore compare Eq.\eqref{eq::gibbs-surface-tension} to those obtained from our DFT minimization, see Tab.\ref{tab:surface_tension}, showing good agreement between both routes of calculating the reduced surface tension where slight differences occur for larger size ratios. As can be inferred from Fig.\ref{fig::total}, the excess adsorption $\Gamma_2$ increases when the size ratio becomes larger. This has in total the effect of lowering the surface tension according to Eq.\eqref{eq::gibbs-surface-tension}.

\subsection{Effective Interactions}\label{sc::effective-interactions}

Finally, the binary SW mixture can also be treated as an effectively 
one-component system of nano particles by integrating out the degrees of 
freedom of the solvent species 
\cite{McMillan45,Dijkstra98,Dijkstra99a,Dijkstra99b,Roth01}. This mapping 
introduces effective interactions \cite{Roth00,Likos01}
in the remaining species of nano particles. Note that in stark contrast to
the models mentioned in the introduction, the ideal colloid-polymer mixture
\cite{AO54,AO58,Lekkerkerker11}
or the hard sphere mixture \cite{Roth00,Ashton11}, where the leading order contribution of the
mapping from a mixture onto an effective one-component system was an
effective two-body interaction, i.e. a pair potential, between particles, here the
presence of the liquid-vapor interface introduces an important and strong 
one-body term. While two- and higher order terms are also a consequence
of the mapping \cite{McMillan45,Dijkstra98,Dijkstra99a,Dijkstra99b,Roth01}, 
we will not consider them here.

The effective one-body term takes the form of a potential $V_\text{eff}(z)$
and is caused by the solvent. In the dilute limit of nano particles, the density
profile $\rho_2(z)$ of the nano particles can be written as
$\rho_2(z)=\rho_{2,0}\exp(-\beta V_\text{eff}(z))$, which in turn implies 
that given the density profile $\rho_2(z)$ stemming from the binary mixture 
the corresponding effective potential has to be
\begin{equation}\label{eq::effective-pot}
    \beta V_\text{eff}(z) = -\lim_{\rho_{2,0}\to 0}\log\left(\frac{\rho_2(z)}{\rho_{2,0}}\right).
\end{equation}
If we start from a true binary mixture, this limit can be obtained 
{\em numerically} by choosing a sufficiently small value of $\rho_{2,0}$.

On the other hand, we can also make use of Widom's insertion theorem
\cite{Widom63,Henderson83,Roth00}, 
stating that the change in grand potential due to inserting a particle is
related to the excess chemical potential of that particle. Here, we consider
the displacement of a particle from the bulk ($z\rightarrow\pm\infty$) towards 
the proximity of the interface, at position $z$, and interpret the associated 
change in grand potential as the effective potential $V_\text{eff}(z)$. In more
detail, the effective potential $V_\text{eff}(z)$ according to the Widom insertion theorem is given by
\begin{equation}\label{eq::wigner-insertion}
    \beta V_\text{eff}(z) = \lim_{\rho_{2,0}\rightarrow 0}\left(c^{(1)}_2(\pm\infty)-c^{(1)}_2(z)\right),
\end{equation}
where we take the dilute limit of vanishing density of the nano
particles. It is interesting to note, that the dilute limit can be taken
explicitly in this route, so that only the density profile of the interface
of the solvent, without the presence of any nano particles, is required. Still,
for the calculation of $c^{(1)}_2$, the one-body direct correlation function of the
nano particles, a theory for a mixture is necessary. Note that $c^{(1)}_2(\pm\infty)=-\beta\mu_{2,\text{ex}}$ as the correct bulk limit of the one-body
direct correlation function $c^{(1)}_2(z)$ at phase coexistence. 

\begin{figure}
\centering
\includegraphics[scale=0.6]{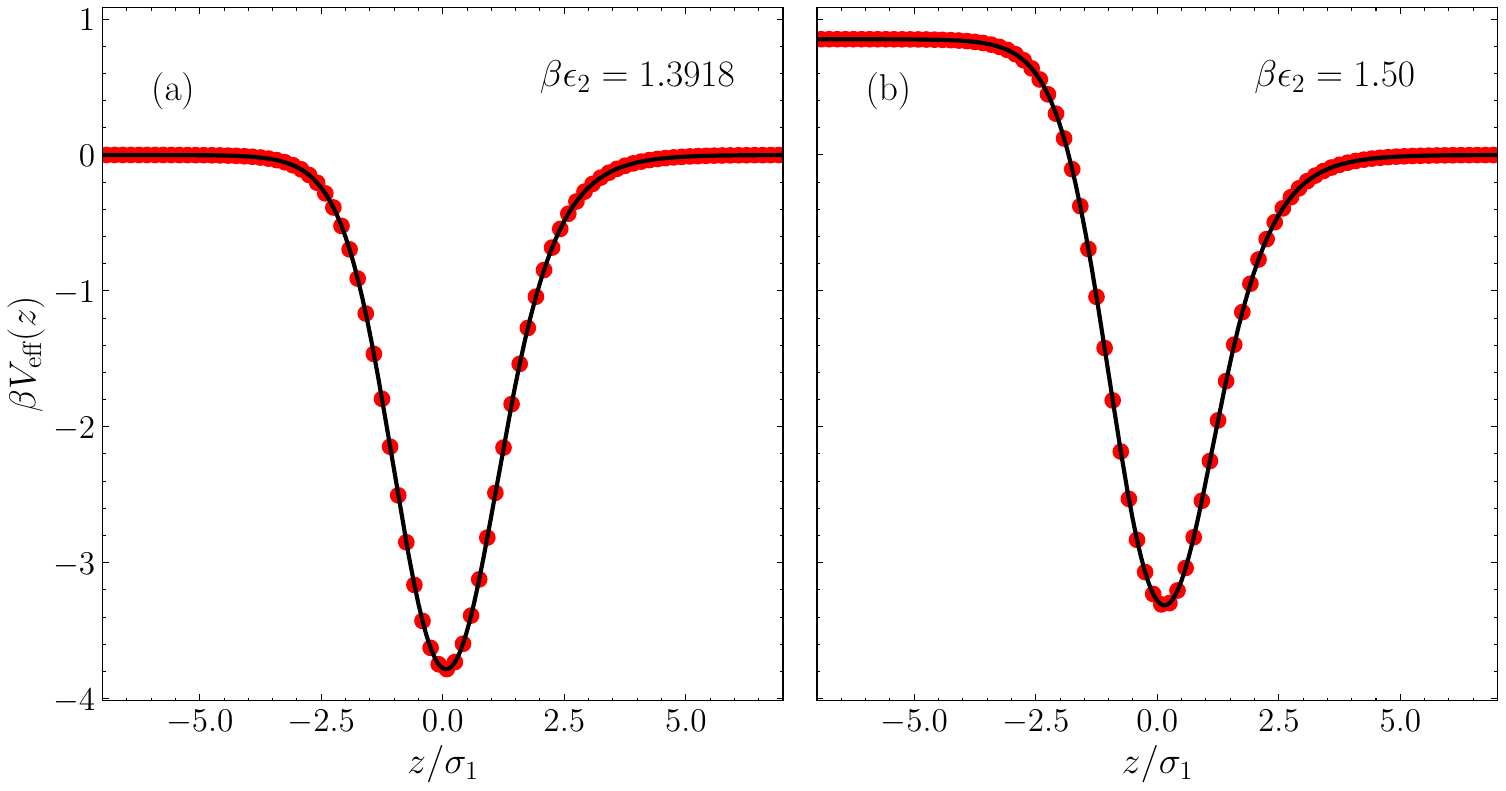}
\caption{Effective potential $V_\text{eff}(z)$ near the interface at $z=0$ once obtained via Eq.\eqref{eq::effective-pot} (red dots) and Eq.\eqref{eq::wigner-insertion} (solid black) of the asymmetric mixture $\sigma_2/\sigma_1=2$ evaluated for the symmetric case 
 $\beta\epsilon_2=1.3918$ (a) and the case $\beta\epsilon_2=1.50$ (b) where the nano particles prefer the liquid. The agreement between both routes is very good.\label{fig::effective-potential}}
\end{figure}

We can compare these two routes to calculate the effective potential, as displayed in Fig.\ref{fig::effective-potential} for one example in the case
$\sigma_2/\sigma_1=2$ with the symmetric solution $\beta\epsilon_2=1.3918$, Fig.\ref{fig::effective-potential}(a), and the solvophilic solution $\beta\epsilon_2=1.50$, Fig.\ref{fig::effective-potential}(b).
Since in the former case the particles do not prefer the liquid to the vapor and 
vice versa, the effective potential $V_\text{eff}(z)$ must tend to 0 when the
bulk either to the left or to the right is reached, which is clearly observable 
in Fig.\ref{fig::effective-potential}(a). Furthermore, we see a minimum at the 
interface which implies that latter is highly attractive for nano 
particles. By slightly increasing the energy $\epsilon_2$ as in Fig.\ref{fig::effective-potential}(b), the nano particles become more solvophilic, i.e. they prefer the liquid to the vapor which in Fig.\ref{fig::effective-potential}(b) is reflected by the fact, that the effective potential becomes positive for $z\rightarrow-\infty$. 

We observe an excellent agreement between the two routes 
of calculating the effective potential which also holds for other size ratios that
we have considered here.

\section{Summary and Outlook} \label{sec:summary}

In this manuscript, we have explored the behavior of a binary square-well 
(SW) mixture within the framework of classical density functional theory 
(DFT). We have extended existing theoretical models for the (optimized)
random phase approximation treatment \cite{Hansen06,Archer17} of the SW perturbation term 
specifically for additive SW mixtures to a new form in terms of weighted
densities that resemble those of fundamental measure theory (FMT) \cite{Rosenfeld89,Roth10}. This
advancement leverages on the geometrical structure inherent in square-well 
interactions and enables predictions of inhomogeneous structures and 
thermodynamic quantities of multi component SW systems in a computationally 
simple form.

An intriguing aspect of our findings is the observation of reduced surface 
tension at liquid-vapor interfaces due to the presence and accumulation of 
nano particles. This phenomenon is particularly pronounced in systems with 
significant size disparities between components, leading to enhanced adsorption
of nano particles at the interface. Similar studies \cite{Sauer2017, Vergara2019} have reported the same behavior when comparing DFT results obtained with WDA to experimental data and Monte Carlo simulations, as well as \cite{Talanquer2001, Okamoto2016} when using RPA. A more closely related analysis to our work is provided by \cite{AbeKoga2014}, which investigates the solubility of a binary mixture and highlights the accumulation of solute particles at the interface for varying interaction energies. In the dilute limit, an intuitive picture of solubility emerges from Wigner's insertion theorem, which introduces a local excess chemical potential. This quantity can be interpreted as the interaction of the solute with solvent particles, averaged over the canonical ensemble.
The adsorption behavior draws a compelling parallel 
with the coffee stain effect \cite{Deegan97}, where non-uniform distribution of particles 
occurs due to differential evaporation dynamics. In our system, the analogous 
effect arises from the interplay between particle size and interaction 
strength, driving the segregation of nano particles to lower the free energy at 
interfaces.

Looking forward, these insights open several promising avenues for future 
research. One intriguing direction is to explore this effective one-component
system further, particularly focusing on the role of higher-order effective 
interaction terms, which could be more complicated due to the influence of wetting or drying \cite{Archer02,Archer05}. Additionally, extending this framework to investigate the effects of external fields or confinement could yield interesting results.

Moreover, the connection between our findings and phenomena such as the coffee
stain effect \cite{Deegan97} suggests potential applications in nanotechnology. For instance,
controlling particle distribution at interfaces could lead to the development
of novel surface coatings or materials with tailored properties. 

\section*{Acknowledgments}
We thank Dr. F. Surfaro for stimulating discussions and helpful comments on the manuscript.

\bibliographystyle{plainnat}
\bibliography{your-bib-file}

\end{document}